\documentclass[]{aa}  
\usepackage{natbib}
\usepackage{graphicx}
\usepackage{txfonts}
\newcommand{\filter}[1]{\mbox{\it #1\/}}              % filter in italics
\newcommand{\micron}{$\mu$m}              % filter in italics
\begin{document}

\title{Scattering from dust in molecular clouds: Constraining the dust grain size distribution  through near-infrared cloudshine and infrared coreshine}
\titlerunning{Constraining the dust grain size distribution in molecular clouds through cloudshine and coreshine}
%   \subtitle{I. Overviewing the $\kappa$-mechanism}

   \author{M. Andersen \inst{1}
     \and
          J. Steinacker   \inst{1}  
   \and W-F. Thi \inst{1} 
   \and L. Pagani \inst{2} 
   \and {A. Bacmann} \inst{1} 
   \and {R. Paladini} \inst{3}}      

   \institute{ (UJF-Grenoble 1/CNRS-INSU, Institut de Plan\'etologie et d'Astrophysique de Grenoble (IPAG) UMR 5274, Grenoble, F-38041, France     \email{morten.andersen@obs.ujf-grenoble.fr }
\and 
LERMA, UMR 8112 du CNRS, Observatoire de Paris, 61, Av. de l'Observatoire, 75014, Paris, France 
\and 
Infrared Processing and Analysis Center, California Institute of Technology, Pasadena, CA 91125, USA}

   %\date{Received September 15, 1996; accepted March 16, 1997}

% \abstract{}{}{}{}{} 
% 5 {} token are mandatory
 
\abstract
{The largest grains (0.5-1 $\mu$m) in the interstellar size distribution are  efficient in scattering near- and mid-infrared radiation. 
 These wavelengths are therefore particularly well suited to probe the still uncertain high-end of the size distribution.
}
{ We investigate the change in appearance of a typical low-mass molecular core from  the \filter{Ks} (2.2 \micron) band to the {\it Spitzer} IRAC 3.6 and 8 micron   bands, and compare with  model calculations, which include variations of the grain size distribution.}
{We combine {\it Spitzer} IRAC and ground-based near-infrared observations to characterize the { scattered light observed  at the near- and mid-infrared wavelengths from}  the core L260. 
Using a spherical symmetric  model core, we perform radiative transfer calculations to study the impact of various dust size distributions on the intensity profiles across the core.  }
{The observed scattered light patterns in the \filter{Ks} and 3.6 \micron\ bands are found to be similar. 
By comparison with  radiative transfer models the two profiles place constraints on the relative abundance of small and large (more than 0.25 \micron) dust grains. 
The scattered light profiles are found to be inconsistent with an interstellar silicate grain distribution extending { only} to 0.25 \micron\ and  large grains are needed to reach the observed fluxes and the flux ratios. { The shape of the \filter{Ks} band surface brightness profile limits the largest grains to 1-1.5 $\mu$m.}
}
{
In addition to  observing coreshine in the {\it Spitzer} IRAC channels, the combination with  ground-based near-infrared 
observations are suited to constrain the properties of large grains in cores.}


   \maketitle
%
%________________________________________________________________

\section{Introduction}

Interstellar dust is a major constituent of the interstellar medium (ISM) and molecular cores. 
Moreover, the dust grains serve as building blocks
for the planets to be formed in the later phases of the star formation process. 
 For the diffuse interstellar medium the dust distribution as a function of size, $a$,  has been determined to be well
approximated by a power-law, $dn/da\propto a^{-3.5}$ { \citep[][known as an MRN distribution]{MRN}} extending to a        
grain size of about 0.25 \micron\ for silicate grains.  
Recently it was found that grains in  low-mass molecular cores scatter mid-infrared (3.6-4.5 \micron) light efficiently 
\citep{steinacker_10,paganisteinacker}, and it was coined coreshine to distinguish it from cloudshine, which is observed at shorter wavelengths from the outer parts of cores \citep{foster,padoan,juvela}.
A radiative transfer analysis based on a dust grain growth model coupled to the cloud density variation showed that it takes grains larger
than typical 0.1 \micron\ ISM grains to scatter mid-infrared radiation \citep{steinacker_10}.
This points towards grain growth already at an early stage of molecular cores.
Grain growth is suggested by several coagulation models with varying efficiency based on the assumptions about turbulence and coagulation  \citep[e.g.][]{ossenkopf,weidenschilling,ormel}. 

There have been independent claims of grain growth in molecular clouds through e.g. mid-infrared extinction measurements \citep[e.g.][]{ascenso,foster13,cambresy} as well as far-infrared/mm thermal emission \citep[e.g.][]{slepnik,roy}. 
Scattering provides an independent and complementary window to study the dust properties in molecular cores and in particular to probe the grain size distribution. 
{ Early work using only the near-infrared showed that indeed large grains would be necessary but the sizes for the larger grains were not determined \citep{lehtinen}. }
Here we show the possibilities to constrain the grain size distribution through near- and mid-infrared observations by combining {\it Spitzer} IRAC data of one particular core, L260-SMM2 (hereafter L260) with archival \filter{J} and \filter{Ks} band data from the VISTA VSH survey. 
The surface brightness profiles are then compared with a radiative transfer model.

L260 is a 
low mass core in Ophiuchus North,  situated out of the Galactic plane (l,b)=(8.6\degr,+22\degr). 
From mm continuum maps, \citet{visser} estimated a total mass of 1 M$_\odot$ for a distance of 160 pc. 
\citet{caselli} suggest a virial mass of 3 M$_\odot$ based on the width of the N$_2$H$^+$ line.  
Although there is a protostellar source west of the core { \citep[Class I, $L_{bol}=0.7$ $L_{\odot}$][]{bontemps}}, the core itself appears starless, without any 24 $\mu$m source and yet shows coreshine.
The core is extended with a relatively low central density of some $n_0=2.2\times10^4$ cm$^{-3}$. 
Line tracers indicate that the core has a very low level of turbulence. \citet{caselli}  found a \mbox{$N_2H^+$} line width of 0.2$\pm$0.02 km\,s$^{-1}$, suggesting that the turbulence contribution to the total linewidth is similar to the thermal width. 
\citet{visser} found no evidence for any outflow activity, which is supported by the narrow line widths. 
There is  thus little evidence for any hidden protostar in the core, which would suggest that the core is relatively young.

The paper is outlined as follows. 
In Sect. 2 we discuss the available data from {\it Spitzer} and the VISTA VSH survey. 
In Sect. 3 we present the immediate  results from  the surface brightness profiles in the different wavelength bands  along a cut through the core. 
Section 4 discusses intensity profiles obtained from a simple core model with varying dust
size distribution. 
Finally we summarize our findings in Sect. 5. 

%__________________________________________________________________

\section{Data}
\subsection{Spitzer data} 
We have utilized the list of cores compiled in \citet{paganisteinacker} and identified L260 as a case where the coreshine is evident  relative to the background and the confusion from field stars is relatively low. 
The core has been observed with {\it Spitzer} as part of the C2D programme \citep{evans} and later during the warm phase of the mission within  the programme ``Hunting Coreshine with {\it Spitzer}'' (PI Paladini). 
The longer wavelength data (8 \micron , \filter{IRAC4}) are obtained from the cold mission while we utilize the 3.6 \micron\ (\filter{IRAC1})  data from the warm mission due to longer integration time and hence better signal to noise. 

We use the recently calibrated data from the archive for the \filter{IRAC4} observations. 
The long wavelength data  were obtained with an exposure time of 12 seconds per frame and 2 frames per position in the mosaic. 
The integration time per pixel is 10.4 seconds and thus the total integration time per pixel is 20.8 seconds. 
The \filter{IRAC1} data were obtained with an exposure time of 30 seconds per position. 
The coverage for almost all the mosaic was 6 frames or more. 
The integration time per pixel is 23.6 seconds and the total integration time per pixel is thus 141.6 seconds, more in the overlap regions.

\subsection{Near-infrared data}
L260 was observed on August 30, 2010 as part of the ESO VISTA VSH survey (PI McMahon) through the filters \filter{J} and \filter{Ks}.
The  on-source integration time is 15 seconds per dither and 2 dithers were obtained on source. 
{ Concatenated frames were obtained as part of the survey with offsets substantially larger than the size of L260. 
We have utilized the pipeline reduced data where the observations in each filter  have  been flat fielded, sky subtracted and co-added. 
There is no evidence for any extended emission  in either the \filter{J} or \filter{Ks} bands { that is not directly related to L260}. }
The data are calibrated to the 2MASS system. 
Zero point calibration was performed for both filters adopting stars in common between the observed dataset and 2MASS.  
Finally, the data were re-binned to the same spatial resolution as the {\it Spitzer} data, i.e from 0.339\arcsec\ to 0.6\arcsec .

\section{Observational Results}
\begin{figure}
  \resizebox{9cm}{!}
  {\includegraphics{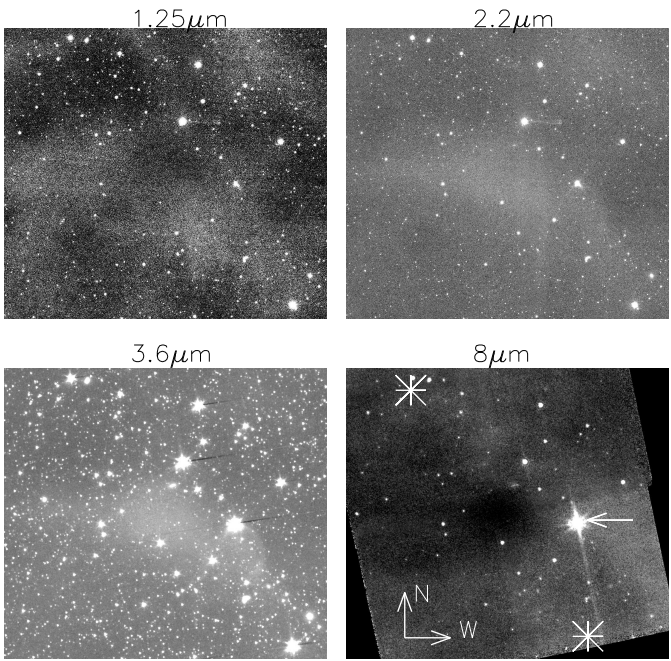}}
  \caption{Top to bottom from left to right: The \filter{J}, \filter{Ks}, \filter{IRAC1}, and \filter{IRAC4} images of L260. {The maps are centred on (RA,DEC)=(16:47:08.35,-09:35:13) and the figures are in equatorial coordinates.}
Marked in \filter{IRAC4} with two asterisks is the extent of the cuts shown in Fig~\ref{L260_cuts} and the arrow indicates the location of the protostar. 
 The field of view of each image is 10\farcm8$\times$9\farcm8.} 
  \label{L260_overview}
  \end{figure}

Figure \ref{L260_overview} shows the image of L260 in the \filter{J}, \filter{Ks}, \filter{IRAC1}, and \filter{IRAC4} filters. 
The centre of the core is easily identified in \filter{IRAC4} where it shows up as an almost circular absorption feature. { There is also indication {that the core has an extension to the east of low column density, traced by a correspondingly low optical depth in the \filter{IRAC4} band.}}

To characterise the scattering from the core as a function of wavelength we have made a  cut through the core at constant Galactic latitude. 
Although the cloudshine and coreshine can readily be seen  in the images in the sense that the {scattered light} follows the extinction observed in \filter{IRAC4}, it is still a relatively faint (some 20\%) phenomenon relative to the background contribution in the images. 
To improve the signal to noise ratio we have therefore binned the pixels perpendicular to the cut direction. 
The average surface brightness over 10 pixels perpendicular to the  direction of the cut   (6\arcsec , or { 960} AU) is used. 
There is no apparent structure in either the cloudshine or coreshine on this scale. 
Further, the flux is smoothed over 5 pixels in the Galactic latitude direction. 
{The random noise for the surface brightness profiles have been estimated from the variation of the pixel values for each average bin.}  
{For all the surface brightness profiles a linear fit has been performed to the background which has then been subtracted. 
The slope is small in both {\it Spitzer} wavelengths. 
However, at the \filter{J} and \filter{Ks} bands the slope is stronger. 
For the \filter{Ks} band the change in background level across the core is 0.1 MJy/sr resulting in a systematic uncertainty on the peak of the surface brightness profile of $\pm$0.05 MJyr/sr. 
For the \filter{IRAC1} band the change in the peak surface brightness is  0.01 MJy/sr.  }

\begin{figure}
  \resizebox{9cm}{!}
  {\includegraphics{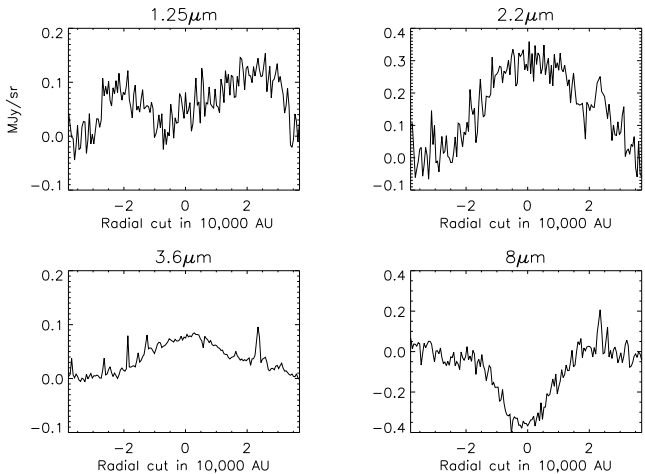}}
  \caption{Surface brightness cuts through the core L260. The location of the cut is shown { in panel 4 in} Fig.~\ref{L260_overview}. The cuts extend along constant Galactic latitude from the right hand side in Figure~\ref{L260_overview} towards the left between the points marked with asterisks. 
The bright spikes most notable in the \filter{IRAC1} profile are due to stars. }
  \label{L260_cuts}
  \end{figure}

The comparison of the surface brightness profiles shown in Fig.~\ref{L260_cuts} illustrates  the difference in morphology of the 3 scattering profiles. 
The \filter{IRAC1} profile is smooth, symmetric, and peaks close to the position with the highest column density as measured in  the \filter{IRAC4} band.
For the \filter{Ks} band surface brightness profile a slight flattening is noticed but only very close  to the centre of the core, as expected if the flattening  is due to extinction effects. 
The \filter{J} band is a bit more complicated. The profile peaks on the Galactic Centre side of the core  and is substantially weaker on the other side due to self-absorption. 
This is due to absorption of the background radiation that is stronger in the \filter{J} band than the scattering of the interstellar radiation field which results in parts of the core appearing in absorption.  
Further, the \filter{J} band profile is complicated by low column density material.  
{ It is likely part of Ophiuchus north since the radial velocity is similar, based on molecular line measurements (Andersen et al. in prep) but there is no evidence that it is directly associated with L260. 
We have chosen not to use it in the modelling. 
In more isolated cores the \filter{J} band is likely to be a further diagnostic of the grain population. }

The scattered light mainly originates from the interstellar radiation field. 
Protostars can emit substantial radiation at near-infrared and mid-infrared wavelengths due to their low temperature, large surface area, and the presence of circumstellar material.
However, the morphology of the scattered light images does not show evidence for a very close by radiation source. 
Indeed, if the protostar located to the west of the core (marked by the arrow in Fig,~\ref{L260_overview}) was strongly illuminating the core it would be  difficult to obtain bright cloudshine, especially in the \filter{J} band, to the east of the core centre which is what is observed. 
On the other hand, the cloudshine and coreshine seen just south of the protostar may have a local origin.

\section{ Scattered light profiles of a model core with the properties of L260}
 So far, only  L183 has been modeled with a detailed 3D structure and dust model to explain
the coreshine in the {\it Spitzer} images \citep{steinacker_10}. 
The dust model assumed a single grain size per location and a power-law relation between particle size and local gas density. 
While \citet{steinacker_10} were able
to fit the observed coreshine morphology, the flux gradients were larger in the model
images than in the observed ones. Moreover, for optically  thin scattering, 
the general good correlation between coreshine pattern
and extinction pattern  for many cores \citep{paganisteinacker} might 
indicate a better mixing of grain sizes throughout the core. Therefore, we test
the  case of a size distribution which is not varying across the core.
 To perform a series of
radiative transfer calculations, we have modified parts of the radiative transfer code 
applied in \cite{steinacker_10} to make use of opacities which are derived from
an integration over the size distribution weighted properties
 \citep[for details, see Sect.~2.4 in][]{steinackerreview}. A spherically symmetric core
with a power-law density profile with index -1.5 and flattening at a radius of 6500 AU was used to illustrate the observables for a core at the location of L260. 
This density profile is able to reproduce the shape of the \filter{IRAC4} surface brightness profile and is in agreement with the parameters provided in \citet{visser}. 

The interstellar radiation field illuminating the core is partly unknown. 
Here, we approximate it by the  anisotropic field as measured by the legacy archive microwave background data analysis (DIRBE/LAMBDA\footnote{http://lambda.gsfc.nasa.gov}) at the Earth location,  rotated to match the core position with respect to the observer. Shadowing effects from surrounding dust clumps have not been taken into account. 
Further diffuse radiation is necessary to reproduce the scattering profile, an additional 1.7  times the general interstellar radiation field is necessary. 
Since L260 is located in a relatively bright star forming region it is to be expected that the near-infrared diffuse radiation is larger than the standard interstellar amount. 
{ The background radiation along the line-of-sight will be attenuated by the core and, in the lack of coreshine, the core would be seen in absorption relative to the background. 
This component therefore has to be characterised. 
For this we have used the DIRBE map since the absolute background is provided as opposed to the {\it Spitzer observations}. 
Due to the large beam of DIRBE the surface brightness in each pixel is contaminated by point sources.  Although the point source contribution should be {added to}  the diffuse radiation, it should not be considered for  the line of sight background absorption. To subtract the flux from point sources we have utilized the WISE point source catalogue. The flux of all objects within the background DIRBE pixel was obtained from the catalogue and subtracted from the DIRBE value. Although the WISE observations saturate at $m_{3.6}=8$ mag, point source photometry of the wings of the saturated stars have enabled photometry of sources up to $m_{3.6}=2$ mag. 

For the three DIRBE pixels closest to L260 the raw DIRBE values are 0.20, 0.23, and 0.18 MJy/sr, respectively. The point source contribution to each was found to be 0.12, 0.10, 0.06 MJy/sr, respectively, providing final star subtracted background estimates of 0.08, 0.13, and 0.12 MJy/sr, respectively. We used the average for the background value. 
The central wavelength of the DIRBE map is very close to the \filter{IRAC1} observations (3.5 vs 3.6 $\mu$m) and the surface brightness between the two wavelengths is expected to be very similar. 
One caveat of the approach is the large beam of DIRBE relative to {\it Spitzer}, 42\arcmin$\times$32\arcmin . 
Depending on the region this can in principle introduce substantial uncertainties if the background at 3.6 $\mu$m varies rapidly on smaller scales than the DIRBE beam. 
However, the fact that the three DIRBE beams overlapping with L260 provides the same surface brightness value after star subtraction suggests this is not the case. 
 A similar approach was adopted for the \filter{Ks} band where the 2.2 \micron\ DIRBE map was used and the stellar contribution to the line of sight was subtracted using the 2MASS point source catalog. 
The DIRBE pixel values at 2.2 \micron are 0.28, 0.32, and 0.23 MJy/sr, respectively and the corresponding point source contributions were 0.23, 0.20, and 0.12 MJy/sr. We use the average value of 0.09 MJyr/sr for the line of sight background at 2.2 \micron\

The dust opacities and phase functions were taken from \cite{drainelee}. 
The grain size distribution was assumed to be an 
MRN type distribution, but with varying upper size limit and different slopes.
Further, the grain composition was modified from astro-silicates to pure silicates, in all cases ice covered. }
{The presence of cloud- and coreshine shows that the scattering of the diffuse interstellar radiation field is stronger than the absorption of the background radiation such that $\Delta I_{obs}=I_{sca}+I_{bg}e^{-\tau}-I_{bg}$  {is positive} \citep[][Steinacker et al. in prep]{lehtinen}. 
The scattering is the product of the interstellar radiation field weighted by the phase function and the albedo of the grains. 
Fig.~\ref{albedo} shows the albedo as a function of grain size for different wavelengths. 
 \begin{figure}
 \includegraphics[width=9cm]{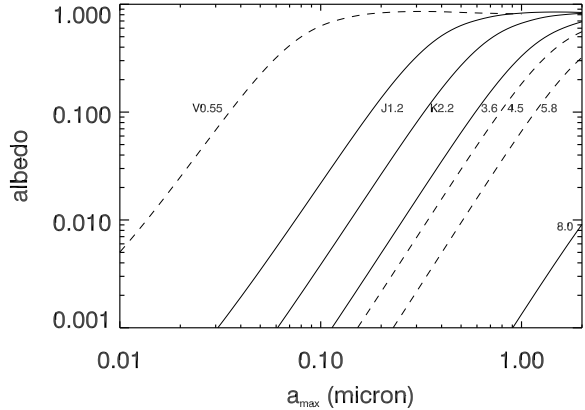}
 \caption{The albedo of the ice-covered silicate dust grains for different wavelengths as a function of the maximum grain size for an MRN distribution. 
For a maximum grain size smaller than 0.8 $\mu$m the albedo is less than 0.2 for  3.6 $\mu$m radiation. 
The dashed lines indicate the albedo for other common filters.}
         \label{albedo}
 \end{figure}
Grains smaller than $\sim1\mu$m have a very small albedo for 3.6 $\mu$m radiation. 
Thus, if small grains are to be responsible for the observed scattering in the \filter{IRAC1} band the radiation field has to be very strong. 
A grain size distribution extending to 0.25 $\mu$m would only have an albedo of 0.01 and the radiation field would have to be increased by a factor of 20 stronger than the interstellar radiation field without altering the background level to compensate for the low albedo. 
There's no evidence for such a strong field which leaves us with the conclusion that large grains are necessary. 

}

\begin{figure}
 \includegraphics[width=9.5cm]{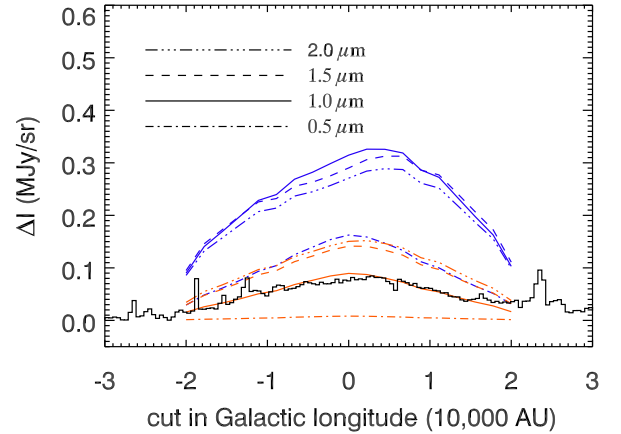}
 \caption{Surface brightness profiles through the centre of a spherical model core with the overall properties of L260 and located in an enhanced diffuse radiation field (1.7 times the interstellar radiation field). 
Shown are the surface brightness profiles in the filters \filter{Ks} (blue) and \filter{IRAC1} (red) under three different maximum sizes { for a MRN type distribution, 0.5 \micron, 1.0 \micron, 1.5 \micron, and 2.0 $\mu$m. A cutoff at 0.5  \micron\ does not produce any coreshine, grains on the order 1 \micron\ are necessary. Dust properties were adopted from \citet{drainelee}. 
The observed 3.6 \micron\ surface brightness profile is shown for reference (black line).}  }

         \label{L260_models}
 \end{figure}

We show in Fig.~\ref{L260_models} the \filter{Ks} and \filter{IRAC1} surface brightness profiles for MRN (slope of -3.5) dust distributions extending to different maximal grain sizes. 
The adopted core mass for the test case is 1.5 M$_\odot$, a compromise between the masses determined from the dust continuum and the dynamical mass, with the density profile described above. 
{
{ For a maximal grain size of 0.5 \micron\  the \filter{Ks} band shows up in emission due to the more effective scattering by large (larger than 0.25 \micron) grains but the grains are not sufficiently large for effective scattering at 3.6 $\mu$m. 
Larger grains are necessary and for grain sizes up to 1 \micron\  there is {coreshine} at 3.6 \micron\ comparable to that observed in molecular cores. 
However, there will be a suppression at the centre of the \filter{Ks} band surface brightness and slightly less scattering on the shadow side of the core due to the higher  extinction at the core centre. 
The shift amounts to 2000 AU for the radiation field adopted for Ophiuchus North and the adopted core shape. 
The higher the column density at the centre the stronger the effect will be. 
This can be seen for the larger cutoffs of 1.5 and 2 \micron . 
The  3.6 \micron surface brightness is too strong simultaneously with the 2.2 \micron surface brightness becoming marginally too weak. 
Since the effects are in opposite directions this cannot be rectified by a simple change in the optical properties of the grains.}
The different behavior between the observed \filter{Ks} and \filter{IRAC1} filters is thus a diagnostic of the relative abundance of small (sufficiently small so that scattering at 2.2 \micron\ is negligible) to large grains and of the largest grain size.

 \begin{figure}
 \includegraphics[width=9.5cm]{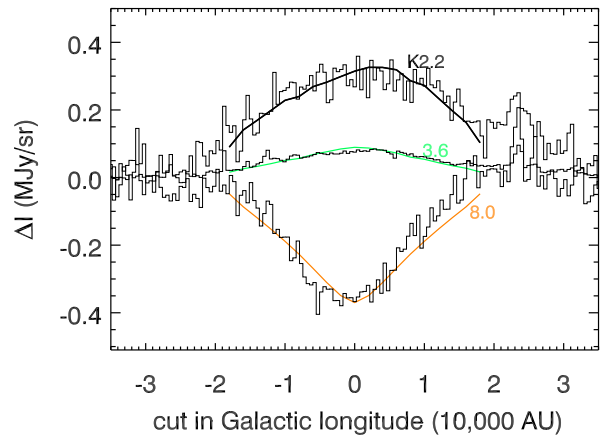}
 \caption{{ A comparison of the surface brightness profiles in the \filter{Ks}, \filter{IRAC1}, and \filter{IRAC4} filters with the corresponding profiles for a model molecular cloud. The observations are shown as black histograms whereas the lines are the model profiles.} 
}
         \label{modelcuts}
 \end{figure}
Fig.~\ref{modelcuts} shows a comparison between the observed surface brightness profiles in the \filter{Ks}, \filter{IRAC1}, and \filter{IRAC4} filters for L260  with the predicted surface brightness profiles for a 1.5 M$_\odot$ model core with a flattening radius of 6500 AU, a maximal grain size of 1.0 \micron, a power-law dust size distribution with a slope of $-3.65$ and a radiation field 1.7 times stronger than the diffuse interstellar radiation field. 
The 1D density model does a reasonable job at reproducing the cuts at different wavelengths across the core. 
We do not attempt a detailed modelling of L260 here, the aim is to show the feasibility of shorter wavelength observations to constrain the dust properties through the combination of extinction and scattering.

Only very few large grains (more than 0.5 \micron) are necessary to produce the observed scattering. 
With the slope adopted only some 4.6$\times 10^{-6}$ of the grains would be larger than 0.5 $\mu$m. 
However, they would contain 25\%\ of the total dust mass.

The low level of turbulence found for the core centre (0.2$\pm$0.02 km\,s$^{-1}$) is puzzling for a scenario where the grains are grown in the core due to coagulation { for typical core life times of 1 Myr or less}. 
All coagulation models rely on the
relative velocities between the grains that are provided by the turbulent motions of the gas they
are kinetically coupled to. 
Moreover, good correlation between the \filter{IRAC1} band coreshine and the \filter{IRAC4} absorption across the core
suggests that there are no strong gradients in the grain size distribution which would speak for
an efficient mixing of locally grown grains. Such mixing becomes difficult with low gas turbulence observed { for the expected life times of molecular cores of only few free fall times (of the order 0.1 Myr) \citep{ward-thompson}}. 
With turbulent velocities of 0.2 km/s \citep{caselli} one core crossing time would be almost 1 Myr and several crossings would be expected to generate a smooth surface density distribution.
Since the growth process is expected to last for a substantial
part of the pre-stellar core lifetime, it is possible that the turbulence was stronger
in the past and decayed during any contraction process. 
An alternative explanation is that  there is no growth and that the grains were present when the core was formed. 
With the MRN distribution determined for the ISM
that has a large-sized grain tail, invisible in extinction measurements, it has to be understood { why coreshine was only detected in 50\%\ of the sources in \citet{paganisteinacker}.}

\section{Conclusions}
A simultaneous modelling of the near- and mid-infrared scattered light  has the potential to constrain the higher-mass end
of dust size distributions
in molecular cloud cores since the scattering seen in  cloudshine and coreshine is very 
sensitive to the largest grains.
Our modelling indicates that it is very difficult to produce sufficient scattered light flux in order
to explain the observed coreshine fluxes when assuming a classical MRN distribution and a maximum grain size of 0.25 $\mu$m. 
Larger grains (up to $\sim$1.0 \micron)   are necessary but they are possibly less abundant than predicted by an extension of the MRN distribution. 
Future modelling will explore the abilities for coagulation models to reproduce the observed coreshine and cloudshine surface brightness levels. 
In the case of the L260, we have demonstrated that low-mass cores are sensitive 
in the \filter{Ks} band to the upper size limit of the size distribution allowing 
a continuous growth analysis  with ground-based telescopes using scattered light. 

\begin{acknowledgements}
{We thank the anonymous referee for many suggestions that improved the manuscript. }
Part of this work was funded by the Agence National Recherche through the Chaire d'Excellence grant ANR (CHEX2011 SEED). 
We thank F.-X. D\'esert for assistance in the processing of the all sky background maps. 
\end{acknowledgements}

\end{document}